\documentclass[a4paper,11pt]{article}
\usepackage{graphicx}
\usepackage{authblk}
\usepackage{amsmath}
\usepackage{xcolor}
\usepackage{subfig}
\usepackage{pos}
\usepackage{hyperref}
\usepackage{url}
\usepackage{comment}

\newcommand{\bs}[1]{\boldsymbol{#1}}
\newcommand{\pd}[2]{\frac{\partial #1}{\partial #2}}
\newcommand{\braket}[1]{\langle #1 \rangle}

\title{NSPT for $O(N)$ non-linear sigma model: the larger $N$ the better}

\author*[a,b]{Paolo Baglioni}
\author[a,b]{Francesco Di Renzo}

\affiliation[a]{University of Parma,\\
  Department of Mathematical, Physical and Computer Sciences, Parco Area delle Scienze 7/A, Parma}

\affiliation[b]{INFN - Gruppo Collegato di Parma,\\
Parco Area delle Scienze 7/A  – Parma}

\emailAdd{paolo.baglioni@unipr.it}
\emailAdd{francesco.direnzo@unipr.it}

\abstract{The $O(N)$ non-linear sigma model (NLSM) is an example of field theory on a target space with nontrivial geometry. One interesting feature of NLSM is asymptotic freedom, which makes perturbative
calculations interesting. Given the successes in Lattice Gauge Theories, Numerical Stochastic Perturbation Theory (NSPT) is a natural candidate for performing high-order computations also in the case of NLSM. However, in low-dimensional systems NSPT is known to display statistical fluctuations substantially increasing for increasing orders. In this work, we explore how for $O(N)$ NLSM this behaviour is strongly dependent on $N$. As largely expected on general grounds, the larger is $N$, the larger is the order at which a NSPT computation can be effectively performed.}

\FullConference{The 40th International Symposium on Lattice Field Theory (Lattice 2023)\\
July 31st - August 4th, 2023\\
Fermi National Accelerator Laboratory\\}

\begin{document}
\maketitle

\section{Introduction to High-Order Fluctuations in NSPT}
Numerical Stochastic Perturbation Theory (NSPT) \cite{DiRenzo1994} is a numerical method for perturbative computations that closely resemble the non-perturbative (Monte Carlo) ones. This allows to effectively deal with perturbation theory, exploiting all the well-known tools of MC simulations (statistical analysis, state-of-the-art algorithms, efficient and optimized hardware) and much more. In addition, a good share of the effectiveness of NSPT is due to its automated order-by-order processing of expressions \cite{DiRenzo2004}, in fact the same of the increasingly popular Automatic Differentiation \cite{Catumba2023}.

Since the first setting up of NSPT, there has been a significant accumulation of expertise, particularly in understanding the underlying stochastic processes. A key observation was that NSPT, when implemented for \textit{small} systems, exhibited significant and irregular fluctuations at orders manageable in \textit{larger} systems \cite{Alfieri2000}. In the former, the properties of the stochastic process are very hard to study, due to the emergence of rare events in the tails of the order-by-order distributions (i.e., one faces the presence of enormous spikes far from the average values). In fact, for low-dimensional models, it is difficult to estimate the perturbative coefficients even at loop orders as low as less than $10$, while for  theories with much more degrees of freedom, like Lattice QCD, loop orders often exceeding $30$ can be achieved \cite{Bali2014, DelDebbio2018}. Recently, this problem has come back into focus as the same fluctuations have been observed in the NSPT perturbative expansions around non-trivial vacua \cite{Baglioni2023}.

This study aims to quantify the observation that large deviations at high perturbative orders correlate with the number of degree of freedom of the model. The $O(N)$ non-linear sigma model (NLSM) serves as an ideal testing ground for this hypothesis, because we can tune the parameter $N$ modifying the number of degrees of freedom.

This is in much the same spirit of \cite{GonzlezArroyo2019}, where a tendency of distributions of NSPT computed coefficients to Gaussian in the large $N$ limit was reported.
\section{\texorpdfstring{$O(N)$}{} NLSM setup for NSPT}
The origins of NSPT go back to the seminal paper by Parisi and Wu on Stochastic Quantization \cite{Parisi1980}. Basically NSPT was introduced as the {\em order-by-order} numerical integration of the Langevin equation (see \cite{DiRenzo2004} for a complete review) 
 \begin{align}
    \label{eq:langevin}
     \frac{d \varphi(\bs{x},\tau)}{d\tau} = -\frac{\partial\mathcal{S}_E[\varphi]}{\partial\varphi(\bs{x},\tau)} + \eta(\bs{x},\tau) 
 \end{align}
Other stochastic equations can be considered as well \cite{DallaBrida2017}. The NSPT program starts taking the field in Eq. \eqref{eq:langevin} and expanding it in a formal power series of the coupling
\begin{align}
\label{eq:pt_exp_field}
    \varphi(\bs{x},\tau) = \varphi^{(0)}(\bs{x},\tau) + \sum_{n=1}^{\infty}g^n \varphi^{(n)}(\bs{x},\tau)
\end{align}
It is clear that, upon inserting this last expression into Eq. \eqref{eq:langevin} and gathering the different orders in the coupling constant, a sequence of partial differential equations is derived
\begin{align}
\label{eq:order_by_order_langevin}
\begin{split}
     \frac{d \varphi^{(0)}(\bs{x},\tau)}{d\tau} & = -G_0^{-1}\varphi^{(0)}(\bs{x},\tau) + \eta(\bs{x},\tau) \\
     \dots\\
      \frac{d \varphi^{(n)}(\bs{x},\tau)}{d\tau} & = -G_0^{-1}\varphi^{(n)}(\bs{x},\tau) + D_n(\varphi^{(0)},\varphi^{(1)}, \dots, \varphi^{(n-1)})
\end{split}
\end{align}
where $G_0^{-1}$ represents the free propagator and $D_n$ the interaction terms coupling various perturbative order of the field. It is important to recognize that the perturbative truncation is exact at any loop order. Given the perturbative solutions of Eq. \eqref{eq:order_by_order_langevin}, any observable can be calculated as follows
\begin{equation}
    O(g,\tau) = O^{(0)}(\varphi^{(0)}) + g O^{(1)}(\varphi^{(0)},\varphi^{(1)}) + g^2 O^{(2)}(\varphi^{(0)},\varphi^{(1)},\varphi^{(2)}) + \dots
\end{equation}
Solving Eq. \eqref{eq:order_by_order_langevin} requires the choice of a numerical integration scheme and of a designated time step $\Delta\tau$. In our study we selected the Euler integrator. This method demands running simulations at several small enough $\Delta\tau$ values, followed by data extrapolation towards a vanishing $\Delta\tau$; this asks for implementing a dedicated $\chi^2$ minimization approach (see \cite{DelDebbio2018} for extensive details).\\

The $O(N)$ non-linear sigma model is an interesting quantum field theory. Different lattice regularization of this model are known. For our purposes, we employ the most basic $2D$ version, expressed as
\begin{align}
    S = -\frac{1}{g}\sum_{x,\mu} \bs{s}_x \cdot \bs{s}_{x+\mu} 
\end{align}
$\bs{s}_x$ representing $N$-component lattice real scalar field constrained by the local condition $\bs{s}_x\cdot \bs{s}_x = 1$. $g$ denotes the coupling constant. Perturbation theory involves removing constraints (we adopt a conventional strategy discussed in \cite{Elitzur1983}). This is done splitting the vector field $\bs{s}_x = (\bs{\pi}_x,\sigma_x)$ and rescaling $\quad \bs{\pi}^2_x \rightarrow g\bs{\pi}^2_x$. In this approach, we obtain the following partition function
\begin{align}
\label{eq:PF_PT}
Z = \int \prod_x\ d\bs{\pi}_x \ e^{-\frac{1}{2}\sum_{x,\mu} \Bigl[ (\Delta_\mu\bs{\pi}_x)^2 -\frac{1}{g}(\Delta_\mu\sqrt{1-g\bs{\pi}_x^2})^2 \Bigr] - \frac{1}{2}\sum_x \log{(1-g\bs{\pi}_x^2)}}
\end{align}
in which the only degrees of freedom are the $\bs{\pi}_x$, which are free of constraints. In this work we consider the energy of the system in perturbation theory: 
\begin{align}
\label{eq:energy}
    E & = -\frac{1}{2V} \pd{\log{Z}}{\Bigl(\frac{1}{g}\Bigr)}= \braket{ \boldsymbol{s}_0\cdot \boldsymbol{s}_1} = g \braket{ \boldsymbol{\pi}_0\cdot \boldsymbol{\pi}_1} + \braket{\sqrt{1 + g\boldsymbol{\pi}^2_0}\sqrt{1 + g\boldsymbol{\pi}^2_1}}
\end{align}
As evident from Eq. \eqref{eq:PF_PT} and Eq. \eqref{eq:energy}, this procedure leads to a cumbersome perturbation theory: for increasing loop order, not only new Feynman diagrams should be considered but also new interaction vertices. In fact only four terms of the perturbative expansion of Eq. \eqref{eq:energy} have been computed \cite{Elitzur1983, Alles1997}. By employing NSPT, it becomes feasible to go beyond the fourth order. In fact NSPT computations are totally unaffected by the diagrammatic perturbation theory complications: they only require implementing Taylor series in order-by-order operations.
\begin{figure}[t]
  \centering
  \includegraphics[width=0.47\linewidth]{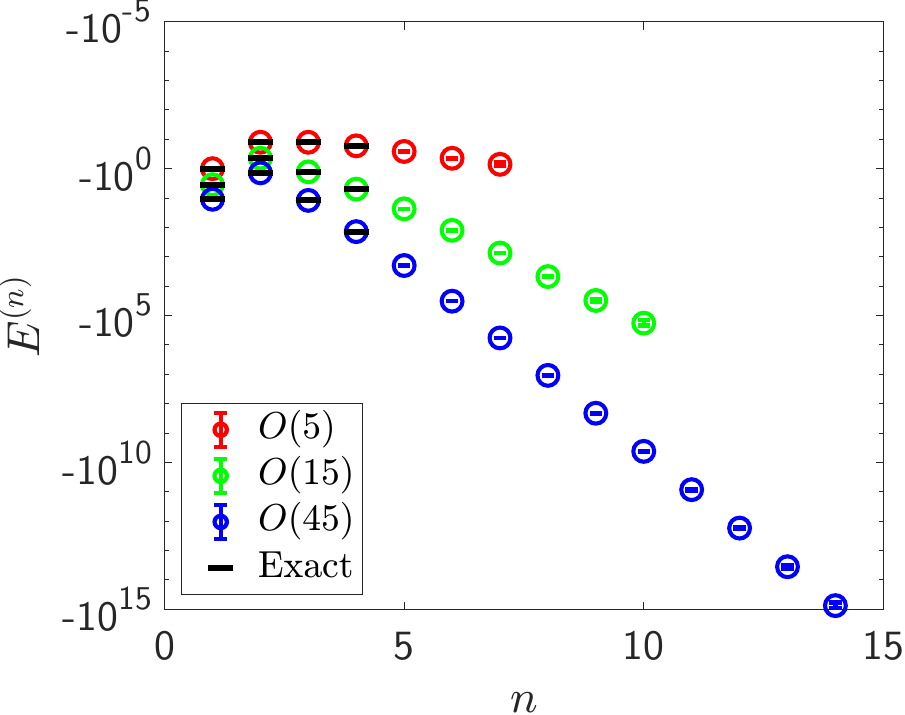}\
  \includegraphics[width=0.47\linewidth]{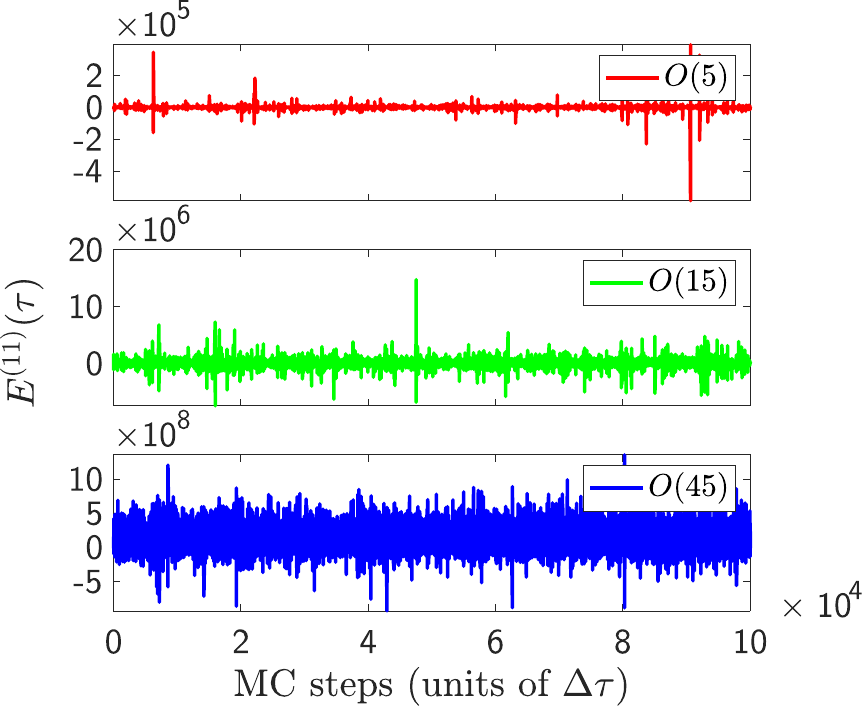}
  \caption{Left plot: There is good agreement between NSPT and known results. We notice that the higher the value of $N$ is, the more NSPT loop corrections we can compute. Right plot: signals from NSPT simulations at different values of $N$.}
  \label{fig:main_and_signal}
\end{figure}
\section{Fluctuations are tamed in the large \texorpdfstring{$N$}{} regime}
We have run a variety of simulations using different $N$ values in the range $(5 - 45)$ on $20\times20$ lattices for $\Delta\tau \in [0.001, 0.01]$. Despite the small lattice size, it turns out that there are very small finite-size effects: as reported in Fig. \ref{fig:main_and_signal} (left plot), we reproduced very accurately the first four known loop coefficients for each value of the parameter $N$. In the case of $N=5$ we pushed the calculation of coefficients up to the seventh loop, while for $N=15$ we were able to estimate coefficients up to the tenth loop. In the case of $N=45$, all coefficients up to the fourteenth loop order have even been estimated. It has been observed (as expected) that for small values of $N$ large fluctuations emerge at high loop orders that completely kill the signal. As said, high-order fluctuations do not allow for a reliable estimation of coefficients beyond order seven (ten) in the case of $N=5$ ($N=15$). To inspect what is going on, we have included in Fig. \ref{fig:main_and_signal} (right plot) an example of NSPT evolution signals for $N=5, 15, 45$ at the perturbative order $n=11$. The signal at small values of $N$ is characterized by very large spikes (typically very rare events but with huge contribution to the mean and standard deviation) that make a consistent determination of coefficient impossible; there is no guarantee that even  simulations of huge duration could make the difference. On the other hand, the signal in the large $N$ limit does not present pathologies.\\
\begin{figure}[t]
  \centering
  \includegraphics[width=0.47\linewidth]{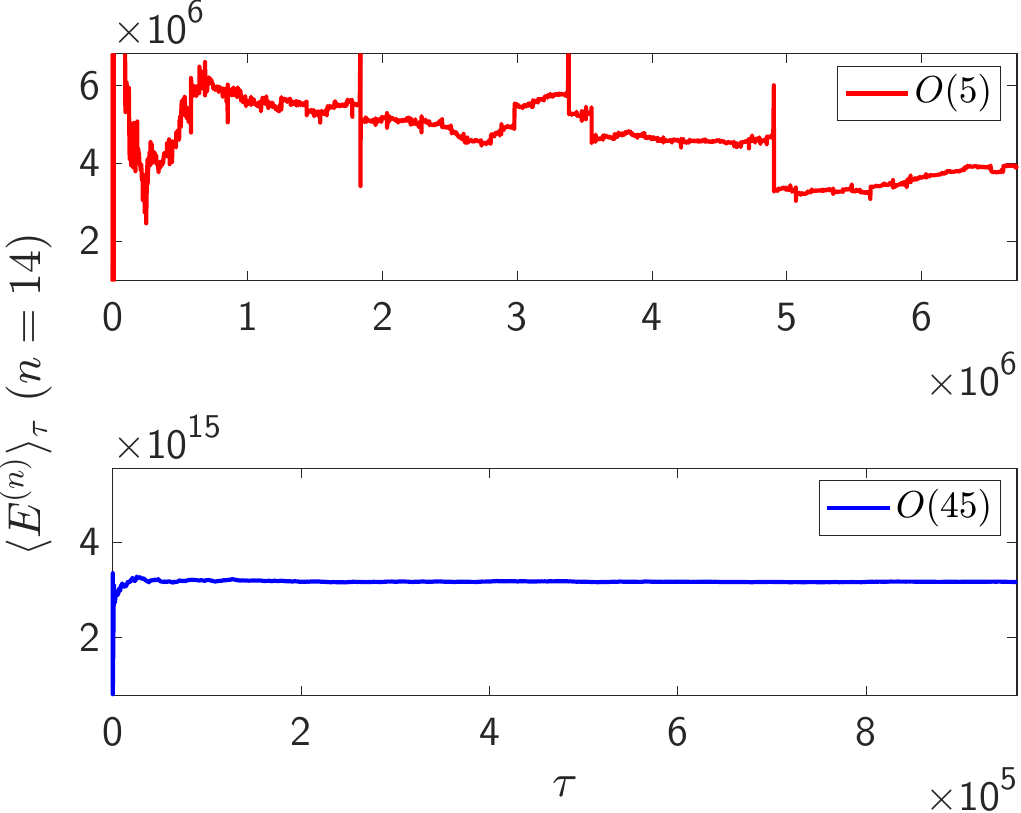}\
  \includegraphics[width=0.47\linewidth]{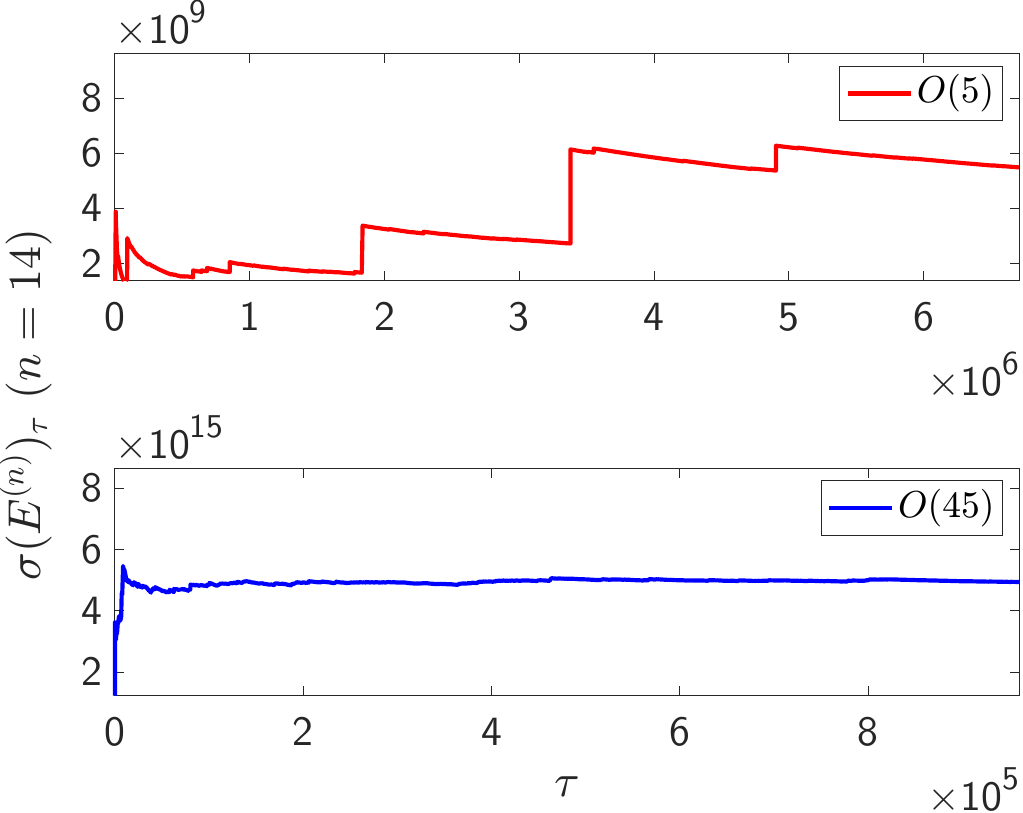}
  \caption{Comparison of the cumulative mean (left plots) and cumulative standard deviation (right plots) for the $O(5)$ and $O(45)$ models. The axes are rescaled to display the same percentage variations. Furthermore, the estimates are made approximately considering the same computational cost.}
  \label{fig:cumsum_and_cumstd}
\end{figure}
In Fig. \ref{fig:cumsum_and_cumstd} we show the cumulative mean and standard deviation, defined respectively as follow
\begin{equation}
    \braket{E^{(n)}}_\tau = \frac{1}{\tau}\sum_{i=1}^{\tau}E^{(n)}_i \quad\quad\quad\quad \sigma(E^{(n)})_\tau = \sqrt{\braket{E^{(n)^2}}_\tau - \braket{E^{(n)}}_\tau^2}
\end{equation}
for $O(5)$ and $O(45)$ at loop order $n=14$. The improvement of NSPT simulations in the large $N$ limit is evident. For $N=5$ there is a great uncertainty in the values of mean and standard deviation (for the latter we cannot even determine if it is finite or not), even considering a huge amount of statistics. In contrast, simulations for the $O(45)$ model are very stable: with the same computational effort which turns out to be insufficient at low values of $N$, we are able to obtain a secure and complete characterization of the NSPT means and standard deviations. All in all, at a given target loop $n^*$, better results are obtained for increasing values of $N$; in particular, there are no issues with the extrapolation in the discrete time step.\\

\begin{figure}[h!]
  \centering
  \includegraphics[width=0.97\linewidth]{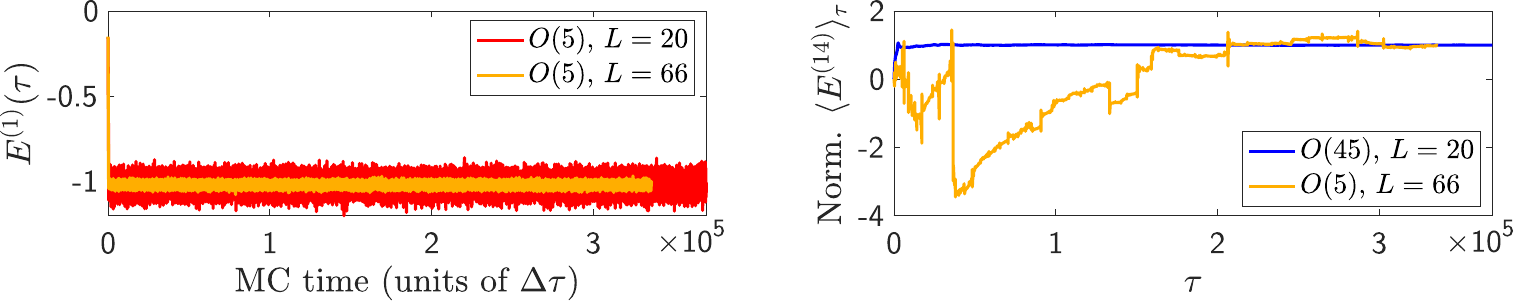}
  \caption{Left plot: simulations of the $O(5)$ model for $L=20, 66$ at low perturbative order. Right plot: simulations at high perturbative orders of $O(5)$ and $O(45)$ on different lattice sizes (this results in roughly the same overall number of degrees of freedom). To compare fluctuations respect to different values, the data have been normalized to have unit mean.}
  \label{fig:diff_L}
\end{figure}

Changing the value of $N$ increases the number of local degrees of freedom; as a matter of fact, also conventional (diagrammatic) perturbation theory is sensitive to that. On the other hand, increasing the size of the lattice $L$ is expected to alleviate some statistical uncertainties through self-averaging effects, but, while there is a large $N$ perturbation theory, there is no large $L$ perturbation theory! In Fig. \ref{fig:diff_L} (left plot) we display the effect of lattice self-averaging at low perturbative order. A large lattice size does not help at all in suppressing large fluctuations at high orders, as reported in Fig. \ref{fig:diff_L} (right plot): here we compute the same perturbative order for different combinations of $N$ and $L$, chosen in such a way that the {\em overall} number of degrees of freedom is the same. The NSPT performance improves only in the large $N$ limit, irrespective of the large $L$ limit: in other words, this is a genuine large $N$ effect (in a sense, $N$ is the only size that matters for NSPT at high order).

Results clearly indicate that fluctuations are tamed in the large $N$ limit, but then a natural question arises: how to estimate up to which perturbative order we can consider a NSPT computation stable for a given $O(N)$ model? Even if we still miss a deeper, detailed understanding of the genesis of fluctuations, it would be useful to have at least a \textit{sanity check}. We provide numerical evidence that the emergence of fluctuations can be signaled through the violation of two simple and robust assumptions on the scaling of relative errors.
\begin{figure}[t]
  \centering
  \includegraphics[width=0.9\linewidth]{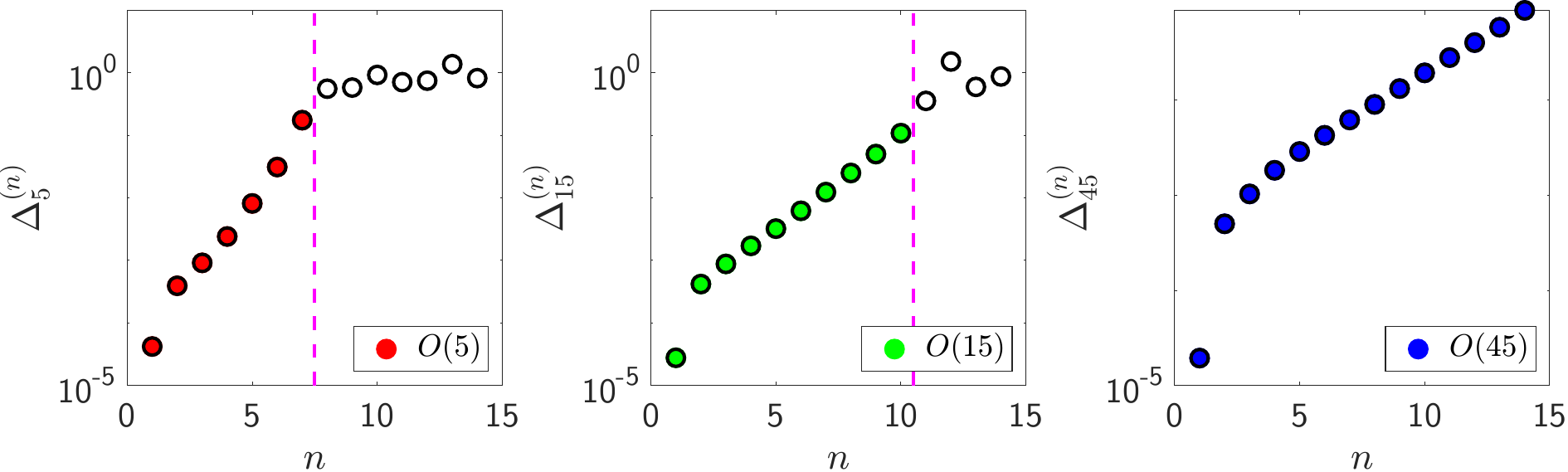}\\
  \includegraphics[width=0.9\linewidth]{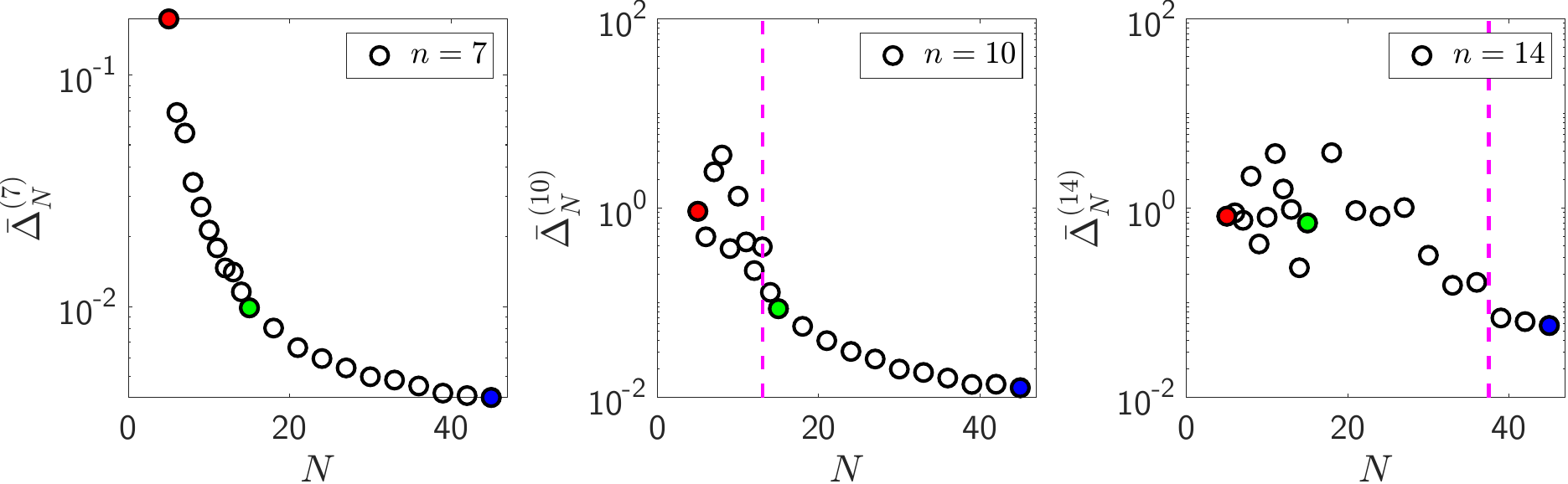}
  \caption{First row: scaling of $\Delta^{(n)}_N$ as a function of the loop order $n$. Second row: scaling of $\bar{\Delta}^{(n)}_N$ at fixed loop order for different values of $N$. Filled marks indicate the points reported in Fig. \ref{fig:main_and_signal}. The dashed magenta lines identify regions beyond which the expected scaling is lost.}
  \label{fig:relativ_error_scaling}
\end{figure}
From now on, we denote relative errors as
\begin{align}
\label{eq:rel_err}
    \Delta^{(n)}_N = \frac{\delta E^{(n)}}{E^{(n)}}\bigg|_{N} \quad \mathrm{and} \quad \bar{\Delta}^{(n)}_N = \frac{\delta E^{(n)}}{E^{(n)}}\bigg|_{N}\cdot \Gamma(N)
\end{align}
where $E^{(n)}$ and $\delta E^{(n)}$ are the extrapolated quantities and the errors at loop order $n$, the symbol $\big|_{N}$ indicates that the quantities are calculated on a fixed $O(N)$ model and $\Gamma(N)$ is a correction factor to compare relative errors at the same statistics. In practice, the factor $\Gamma(N)$ is used to normalize the relative error at the same number of samples. \\
Our first assumption is that the first quantity in Eq. \eqref{eq:rel_err}, plotted as a function of the loop order $n$, should exhibit a monotonically increasing trend. As shown in Fig. \ref{fig:relativ_error_scaling} (first row), deviations from this hypothesis are not observed for the $O(45)$ model up to loop order $n=14$. In contrast, relative errors for the $O(5)$ and $O(15)$ models show a smooth behavior up to a given order, beyond which indeed they violate the hypothesis. We can also study the scaling of the second quantity in Eq. \eqref{eq:rel_err} as a function of the parameter $N$ at fixed $n$. Also for this scaling there is a quite natural hypothesis: as $N$ increases, we are considering systems with an increasing number degrees of freedom, but at the same statistics, and so on very general ground it is expected that $\bar{\Delta}^n_N$ should exhibit a monotonically decreasing trend. Violations of this second hypothesis at a given loop order offers a preliminary indication of a possible $N$ region beyond which NSPT simulations are not contaminated by fluctuations. An example is shown in Fig. \ref{fig:relativ_error_scaling} (second row): in that case, for $n=10, 14$, a validity region for the hypothesis was identified. If we compare the color codes, it is noteworthy that the violations of the two hypotheses in Fig. \ref{fig:relativ_error_scaling} are consistent with each other, ultimately motivating the main plot in Fig. \ref{fig:main_and_signal}. For what has been studied  so far, for each target loop order, a sufficiently large $N$ can always be found, where fluctuations do not yet emerge at that order.
\section{Conclusions}
Despite NSPT suffers from large fluctuations in low-dimensional models, we have numerically demonstrated that for the $O(N)$ NSLM these fluctuations are strongly alleviated in the large $N$ limit. We have provided a rough estimate of what \textit{large $N$} means in the context of NSPT simulations, as we have numerically shown that, at fixed loop order, the expected scaling of the relative errors is restored for values of $N$ larger than a certain $N^*$. In this sense, we can say NSPT is fit for perturbative computations in $O(N)$ NSLM . Interestingly, this class of models is very similar to $CP^{N-1}$ models, which are of great interest because they contain instantons \cite{zakrzewski1989}, around which we can play the same game of \cite{Baglioni2023}. Moreover, this work opens the doors to new asymptotic studies in low-dimensional models. Among these, an interesting one is certainly the study of renormalons: as of now, simulations are ongoing and preliminary results appear to confirm this further possibility.

\section{Acknowledgments}

This work was supported by the European Union Horizon 2020 research and innovation programme under the Marie Sk\l odowska-Curie grant agreement No 813942 (EuroPLEx) and by the INFN under the research project (\textit{iniziativa specifica}) QCDLAT. This research benefits from the HPC (High Performance Computing) facility of the University of Parma, Italy.

\bibliographystyle{JHEP}
\bibliography{reference}

\end{document}